\documentclass[11pt, oneside]{article}   	

\usepackage{graphicx}  
\usepackage{dcolumn}   
\usepackage{bm}        
\usepackage{amssymb}   
\usepackage[parfill]{parskip}    
\usepackage{color}		

\def\d{\delta}
\def\uv{{\mathbf{u}}}
\def\rv{{\mathbf{r}}}
\def\blp{\bigg(}
\def\brp{\bigg)}
\def\la{\langle}
\def\ra{\rangle}
\def\l{\lambda}
\def\F{{\cal{F}}}
\def\Dr{{\Delta\rv}}
\def\H{{\cal{H}}}
\def\D{\Delta}
\def\Q{{\mathbf{Q}}}
\def\P{{\mathbf{P}}}

\def\w{\omega}

\begin{document}

\title{Scaling regimes for wormlike chains confined to cylindrical surfaces under tension}
\author{Greg Morrison$^{1,2}$ and D. Thirumalai$^{3,4}$\\
1:  Department of Physics, University of Houston, Houston TX 77204\\ 
2:  Center for Theoretical Biological Physics, Rice University, Houston TX 77005\\
email:{gcmorris@central.uh.edu}\\
3:  Departments of Chemistry,The University of Texas at Austin, Austin, TX, 78712\\
4:  Department of Physics,The University of Texas at Austin, Austin, TX, 78712\\
email:{dave.thirumalai@gmail.com}
}

\date{\today}

\maketitle

\begin{abstract}
We compute the free energy of confinement $\F$ for a wormlike chain (WLC), with persistence length $l_p$, that is confined to the surface of a cylinder of radius $R$ under an external tension $f$ using a mean field variational approach.  For long chains, we analytically determine the behavior of the chain in a variety of regimes, which are demarcated by the interplay of $l_p$,  the Odijk deflection length ($l_d=(R^2l_p)^{1/3}$), and the Pincus length ($l_f = {k_BT}/{f}$, with $k_BT$ being the thermal energy).  The theory accurately reproduces the  Odijk scaling  for strongly confined chains at $f=0$, with $\F\sim Ll_p^{-1/3}R^{-2/3}$. For moderate values of $f$, the Odijk scaling is discernible only when ${l_p}\gg R$  for strongly confined chains.  Confinement does not significantly alter the scaling of the mean extension for sufficiently high tension. The theory is used to estimate unwrapping  forces for DNA from nucleosomes.    
\end{abstract}


\begin{figure}[t]
\begin{center}
\includegraphics[width=.5\textwidth]{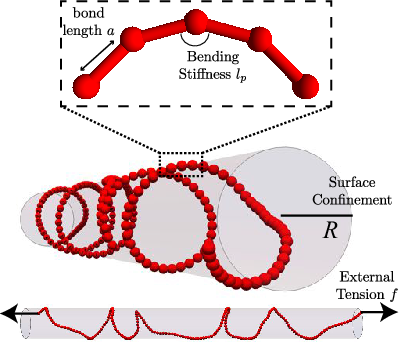}
\caption{Schematic diagram of the chain backbone, with a fixed distance between monomers $a$ and resistance to bending characterized by the persistence length $l_p$, the confinement to the surface of a cylinder of radius $R$, and an external tension $f$ acting on the endpoints of the chain.}
\label{schematicFig.fig}
\end{center}
\end{figure}

\section{Introduction}

There are compelling reasons for understanding the static and dynamics of confined polymers because of their relevance in filtration, gel permeation chromatography, translocation of polymers and polypeptide chains through microporous membranes, and passage of newly synthesized proteins through the ribosome tunnel. These and other considerations prompted several theoretical studies, starting with pioneering studies  \cite{Casassa69Macromolecules,Daoud77JDP}, which triggered subsequent theories and simulations that probed the fate of flexible polymers in pores with regular geometries \cite{Brochard-Wyart90Macromolecules,Morrison05JCP,Jun08PRL,Micheletti11PhysRep,Jung12SoftMatter,Werner14PRE}  as well as in the related case of random media \cite{Edwards88JCP,Thirumalai88PRA} is well understood.

The situation is somewhat more complicated when considering semi-flexible polymers or worm-like chains, WLC, in confined spaces.   Spatial confinement of WLC plays an important role in many biological systems,  \cite{Amitai17PhysRep,Chen16ProgPolymSci,Reisner12dna,Ha15SoftMatter,Saleh15JCP,Spakowitz05BJ}, including histone wrapping in chromatin \cite{Langowski,Schiessel,Cui00PNAS,BenninkNatStructBiol01, Wang11Macromolecules} and nanolithography \cite{Reisner12dna,OdijkConfirmedExpt}.  Here, we consider a WLC that is wrapped around a cylinder whose radius is $R$, and is subject to mechanical force (Fig. \ref{schematicFig.fig}).   Besides the total contour length ($L$), there are three  length scales that control the statistics of the WLC polymer.  The first is the persistence length $l_p$, which is a measure of the resistance of polymers to bending. For long unconfined chains the free energy is a function of $l_p$.  In the mean field theory, proposed here,  the polymer is globally restricted to be wound around the cylinder, which is equivalent to restraint enforced by a soft harmonic potential. Consequently, the Odijk length or  the deflection length, $l_d=(R^2l_p)^{1/3}$, emerges \cite{odijk} coupling the chain stiffness and radius of confinement.  In many biological contexts, the system is often under an external field elongating the chain,  such as  external tension ($f$) unravelling histone-wrapped DNA for replication \cite{BenninkNatStructBiol01}.  An external tension or mechanical force is captured by the Pincus length \cite{Pincus76Macromolecules}, $l_f =(\beta f)^{-1}$ with $\beta=1/k_B T$.  The interplay of $l_f$, and $l_p$, and $l_d$ on the conformations of the WLC is not fully understood. The problem has some relevance to nucleosomes, consisting of DNA wrapped around histone proteins, which is a building block for chromosomes. {\textcolor{black}{This theory may also be applicable to the aggregation of charged biomolecules to surfaces\cite{adsorption1,adsorption2}, which lack the specific interactions that give rise to structure or the nucleosome.}}  Hence, the approximate theory developed here might illustrate an aspect of polymer theory in describing the physics of chromatin \cite{Amitai17PhysRep,Schiessel}.

In this paper, we propose a mean-field approach to study the properties of a wormlike chain confined to the surface of a cylinder \cite{Yang07PRE} under the application of an external tension. Because $l_p$ is comparable to the contour length, $L$, excluded volume interactions may be neglected. We recover the  known \cite{odijk,OdijkConfirmedSim,OdijkConfirmedExpt,odijkHarmonic} dependence of free energy on the deflection length $l_d=(R^2l_p)^{1/3}$ at $f=0$. The theory predicts  the coefficient of the leading term of the scaling of free energy, which is in good agreement with numerical results \cite{Yang07PRE}.  For moderate values of $f \ne 0$ (possibly relevant to DNA unwrapping from nucleosomes) and strong  confinement (${l_p}/{R} \gg 1$), we show that that the free energy scales quadratically with $l_f$.  At high external tensions, we find that the effect of confinement is perturbative, with $1-\langle Z\rangle/L\sim \sqrt{{l_f}/{l_p}}$, independent of the radius of confinement.
%
%

\section{Mean Field Theory}

\subsection{Formulation and approximations}

In order to understand the equilibrium properties of a cylindrically confined WLC under tension, we developed a mean-field theory to arrive at analytically tractable results.  The energy of a configuration of the chain is determined by a variety of terms that are sketched in Fig. \ref{schematicFig.fig}. The system is characterized by  the spacing between monomers ($a$), the number of bonds in the chain (the chain length $L=(N-1)a \approx Na$ with $N \gg 1$), the persistence length of the unconfined chain ($l_p$), the confinement radius ($R$) aligned with the $z$-axis, and the external tension ($f$) that is also applied in   the $z$ direction.  Let the position of each monomer be $\rv_i=(x_i,y_i,z_i)$, and define $\uv_i=\rv_{i+1}-\rv_i$ as the bond vector, and $\hat\uv_i=\uv_i/|\uv_i|$ the unit bond vector.  The statistics of a surface-confined chain can be described by a constrained Kratky-Porod (KP) Hamiltonian \cite{DoiEdwardsBook} $\beta H_{KP}=\frac{l_p}{a}\sum_{n=0}^N \hat\mathbf{u}_i\cdot\hat\mathbf{u}_{i+1}-\beta f (z_N-z_0)$, with confinement to the surface of the cylinder requiring $x_i^2+y_i^2=R^2$ for all $i$.

The KP model is mathematically difficult to work with because of two rigid constraints:  the fixed bond length, $|\uv_i|=a$, and the constraint that monomers be spatially constrained transverse to the $z$ axis, which should be enforced through the relation, $x_i^2+y_i^2=R^2$.  On the mean field level, we replace these rigid constraints for the confined WLC with softer harmonic restraints (an approach that has been fruitfully applied previously \cite{Morrison09PRE,odijkHarmonic,MFforce,Wang07JMS,newMFPaper}).  The form of the MF Hamiltonian  can be found by writing the monomer distribution function explicitly, in order to identify a physically meaningful harmonic approximation to the rigid constraints.  It is straightforward to show that (up to a constant) the statistical weight is
$
\Psi_S=(\prod_{n=0}^N\d[x_n^2+y_n^2-R^2])\times (\prod_{n=1}^{N}\d[{|\Dr_n|-a}])\times (\prod_{n=1}^{N-2}e^{-l_p/2a^3(\Dr_{n+1}-\Dr_n)^2})\times e^{+\beta f(z_L-z_0)}
$
where $\Dr=\rv_{n+1}-\rv_{n}$ and $\beta=1/k_BT$.  The first term enforces confinement of the chain to the cylinder (affecting only the $x$ and $y$ coordinates of the polymer), the second term enforces the constant monomer spacing constraint, the third term accounts for the WLC's resistance to bending, and the fourth term accounts for the external tension.  Each of the $\delta$ functions may be written as $\delta(x_n^2+y_n^2-R^2)\propto \int dK_n e^{iaK_n[(x_n^2+y_n^2)/R^2-1]}$ and $\delta(|\Dr|-a)\propto \int d\Lambda_n e^{-ia\Lambda_n[\Dr^2_n/a^2-1]}$, leading to,
\begin{eqnarray}
\Psi_S&\propto& \int_{-i\infty}^{i\infty}\prod_n dk_n d\l_n\exp\bigg[-a \sum_{n=1}^N k_n\blp\frac{x_n^2+y_n^2}{R^2}-1\brp\label{BigExact}\\
&&-a\sum_{n=1}^{N-1}\l_n\blp\frac{\Dr_n^2}{a^2}-1\brp-\beta f(z_L-z_0)-\frac{1}{2}a \sum_{n=1}^{N-2}l_p\frac{(\Dr_{n+1}-\Dr_n)^2}{a^4}\bigg].\nonumber
\end{eqnarray}
The $\delta$ functions constrain enforce a fixed monomer separation and fixed transverse distance in the confined dimension.  Here, we  focus solely on surface confinement, but note that a MF approach has been applied to the confinement to the interior and surface \cite{Morrison09PRE} of a sphere. An equivalent volume constraint could be applied the interior of the cylinder through a constraint of $x_i^2+y_i^2\le R^2$.  However,  a harmonic constraint on volume confinement requires an estimate of the average distance of each monomer in the transverse direction (which cannot be predicted at the mean field level).  

Although exact, the formulation in Eq. \ref{BigExact} is difficult to work with directly.  Analytical progress becomes possible by assuming that the integrals over the Fourier variables are sharply peaked, in the same manner as in our previous studies \cite{Morrison09PRE,davePaper}.   The partition function $Z=\prod_n d^3\rv_n \Psi_S(\{\rv_n\})\equiv\int\prod_n d\l_n dk_n\exp(-\F[\{\l_n,k_n\}])$ defining the nondimensional free energy functional ${\F}$ as an integral over all the monomer coordinates, and the linearity of the Fourier transform allows us to write $\F=\F_x+\F_y+\F_z-a\sum_{n=1}^N k_n-a\sum_{n=1}^{N-1}\l_n$, with $e^{-\F_x}\equiv\int\prod_n dx_n e^{-\H_c[\{x_n\}]}$, $e^{-\F_y}\equiv\int\prod_n dy_n e^{-\H_c[\{y_n\}]}$,  and $e^{-\F_z}\equiv\int\prod_n dz_n e^{-\H_u[\{z_n\}]}$, where we define the confined Hamiltonian as, 
\begin{equation}
\H_c[\{x_n\}]\equiv a\sum_{n=1}^{N-2}{l_p(\D x_{n+1}-\D x_n)^2}/{2a^4}
+a\sum_{n=1}^{N-1}\l_n{\D x^2_n}/{a^2}+a\sum_{n=1}^{N}k_n{x_n^2}/{R^2}\label{confinedDiscrete}.
\end{equation}
 The unconfined Hamiltonian is,  $\H_u[\{z_n\}]\equiv a\sum_{n=1}^{N-2}{l_p(\D z_{n+1}-\D z_n)^2}/{2a^4}+a\sum_{n=1}^{N-1}\l_n{\D z^2_n}/{a^2}+\beta f \sum_{n=1}^{N-1}\Delta z_n.$  If the free energy is sharply peaked around some $\{\l_n,k_n\}=\{\l_n^*,k_n^*\}$ that minimizes $\F$, the partition function can be written approximately as $Z\approx Z^*=e^{-\F^*}$.  Because the Hamiltonians $\H_c$ and $\H_u$ are uncoupled and quadratic in the monomer coordinates, it is straightforward to integrate over the internal coordinates of the polymer exactly.  Along the confined axes, we find $\F_x=\F_y=\frac{1}{2}\log[{\mbox{Det}}(\Q)]$, where the elements of the matrix $(\Q)_{nm}$ are the coefficients associated with  $x_nx_m$ in Eq. \ref{confinedDiscrete}. The explicit form of $(\Q)_{nm}$ is given in Appendix A of an earlier wor k  \cite{Morrison09PRE}.   In the unconfined direction, completing the square and noting the translational invariance of the system along the cylinder axis,  allows us to write $\F_z=\frac{1}{2}\log[{\mbox{Det}}(\P)]e^{\frac{1}{4}(\beta f)^2 a\sum_{n=1}^{N-1}\l_n^{-1}}$, with \cite{davePaper} $(\P)_{ij}=\l_i\d_{ij}-l_p\d_{i,j\pm1}/2a^2$.  These expressions can in principle be used to determine the stationary phase values for the Fourier variables by setting $\partial \F/\partial \l_n|_{\{\l_n,k_n\}=\{\l_n^*,k_n^*\}}$ = $\partial \F/\partial k_n|_{\{\l_n,k_n\}=\{\l_n^*,k_n^*\}}=0$.  However, the resulting $2N-1$ equations become intractable for large $N$, thus making it necessary to make additional approximations.

The matrices $\P$ and $\Q$ are bidiagonal and tridiagonal, respectively, with  a regular structure except near the endpoints of the chain.  The high symmetry of the matrices that underly the equations suggest that we  should seek symmetric solutions for the stationary values of $\l_n^*$ and $k_n^*$.  In both the unconfined\cite{davePaper} and spherically confined\cite{Morrison09PRE} cases, it was shown that the tractable equations that reproduced exact theoretical results could be found for mean field parameters separated into bulk terms and endpoint terms\cite{davePaper,Morrison09PRE}, with $\l_n^*=\l$ and $k_n=k$ for the interior points on the chain (those with $2<n<N-2$).  Excess endpoint fluctuations generally require that $\l_1=\l_{N-1}\ne \lambda$ and $k_1=k_N\ne k\ne k_{2}=k_{N-1}$ (with the equalities due to symmetry arguments).  This approximation allows us to write $\H_{c}[\{x_n\}]=\H_{c}^{(b)}[\{x_n\}]+\H_{c}^{(e)}[\{x_n\}]$ and $\H_{u}[\{\Delta z_n\}]=\H_{c}^{(b)}[\{\Delta z_n\}]+\H_{c}^{(e)}[\{\Delta z_n\}]$, where the superscript $b$ denotes an extensive bulk term, depending on the mean field parameters $\l$ and $k$. The superscript $e$ denoting an intensive endpoint term depending on the more complicated values of the mean field variables near the ends of the chain. 

In the continuum limit, it can be shown that the bulk form of the Hamiltonians become,
\begin{eqnarray}
\H_c^{(b)}[x(s)]&=&\int_0^L ds \blp\frac{l_p}{2}\ddot x^2(s)+\l\dot x^2(s)+k\frac{x^2(s)}{R^2}\brp\label{bulkConfined}\\
\H_u^{(b)}[\dot z(s)]&=&\int_0^L ds \blp \frac{l_p}{2}\ddot z^2(s)+\l \dot z^2(s)-\beta f \dot z(s)\brp.\label{bulkUnconfined}
\end{eqnarray}
The form for $\H_c^{(b)}$ is identical to  the spherically confined Hamiltonian of the wormlike chain in our previous work \cite{Morrison09PRE}, while the form of $\H_u^{(b)}$ is identical to that for an unconfined chain \cite{davePaper}.  Note that $\l$ is the same in the two Hamiltonians in Eq. \ref{bulkUnconfined} because of  the condition $\la \uv^2\ra=\la u_x^2+u_y^2+u_z^2\ra=1$ couples the three components of the bending vector.   However, the mean field parameter $k$ occurs only in the confined Hamiltonian and enforces $\langle x^2+y^2\rangle=R^2$ constraint.  The assumption that  $\lambda$ is isotropic causes inaccuracies in the predicted mean extension of a chain using the MF approach \cite{MikesPaper}. A MF theory that avoids this assumption has been developed \cite{newMFPaper}, which does  at the cost of greater complexity in the model.  As we will show below, the expected scaling of the free energy in the limits of high tension and strong confinement are both recovered using an isotropic assumption, suggesting that the overall scaling laws will be accurate but with potentially inaccurate coefficients.    We also ignore the endpoint effects by neglecting the Hamiltonians $\H_c^{(e)}$ and $\H_u^{(e)}$.  This approximation simplifies the mathematics greatly, but restricts our analysis to very long chains.  Neglecting the endpoint effects, the mean field equations for the extensive contribution to the free energy in the continuum are $\partial \F/\partial \l=\partial \F/\partial k=0$, with $e^{-\F}=\int{{{\cal{D}}}}[\rv(s)] e^{-\H_c^{(b)}[x(s)]-\H_c^{(b)}[y(s)]-\H_u^{(b)}[z(s)]+L(\l+k)}$.  

{\textcolor{black}{
We note that this model assumes the tension is applied only in the $z$ direction.  In many contexts (including chromatin\cite{Langowski,BenninkNatStructBiol01,Cui00PNAS,Schiessel}), pulling forces may be applied in the transverse ($f_\perp$) and longitudinal ($f_{||}$) directions, with only the latter aligned with the cylinder axis.  Due to the constraints on the system, the component of the force along the cylinder axis will dominate the contribution of the force to the free energy, as the chain is free to elongate in that dimension (making its contribution extensive).  Forces applied in the transverse direction ($f_\perp$) will alter the behavior of the endpoints of the WLC, but will not have a significant affect on the free energy scaling..  We expect these theoretical results will be applicable for WLC's bound to the surface of a cylinder so long as $L/l_p\gg 1$.  The relevant tension scale will be $f_{||}$, the component of the force applied along the axis of the cylinder.  }

{\textcolor{black}{
An analytical advantage of the MF approach is that there is no coupling between the confined x-y dimensions and the tensile z-dimension.  This approximation should break down for sufficiently large forces\cite{MikeTensionPaper}, but the MF approach does accurately recover the statistics confined chains in the absence of the tension, and the scaling of the extension for chains under tension in the absence of confinement.  We note that while these scaling laws are correctly predicted,  the scaling coefficients are not accurately determined using the MF method.  We, therefore, expect the results in this paper to be accurate in the scaling laws predicted, but cannot quantitatively predict the free energy or extension of a confined WLC under tension.
}}
{\textcolor{black}{
The MF theory is analytically tractable in the limit of large $L$, where $\sinh(L/l_p)\approx\cosh(L/l_p)\approx e^{L/l_p}/2$.  The analytic predictions are expected to be accurate so long as the chain is free to slide on the surface of the cylinder.  
}}

\subsection{Calculation of the Free Energy}

It is straightforward to perform the path integrals over the confined \cite{Morrison09PRE} and unconfined \cite{davePaper} dimensions to explicitly compute the partition function $e^{-\F}$, from which it is possible to derive the mean field equations for constant $\l$ and $k$.  One can readily recognize that Eq. \ref{bulkUnconfined} describes a quantum harmonic oscillator after a change of variables, for which an exact propagator is known\cite{FeynmanStatMech}.   
In the confined dimension, it is straightforward to show\cite{Morrison09PRE} that the action in Eq. \ref{bulkConfined} is minimized by a path satisfying $\frac{l_p}{2}x^{(4)}(s)-\l\ddot x(s)+\frac{k}{R^2}x(s)=0$.  It is readily observed (after some tedious mathematics) that the solution is expressible in terms of the frequencies $\w_{\pm}^2=\frac{\l}{l_p}(1\pm\sqrt{1-{2kl_p}/{\l^2R^2}}\ )$.  It is possible to integrate over the internal degrees of freedom of the chain exactly \cite{Morrison09PRE}, which results in  an unwieldy expression.  However, with the assumption that $\sinh(L\w_{\pm})\approx \cosh(L\w_{\pm})\approx e^{L\w_{\pm}}/2$ (satisfied for $L/l_p\gg 1$), it is possible to simplify the confinement free energy as, 
\begin{eqnarray}
\frac{\F}{L} \approx \bigg(\w_++\w_--k\bigg)+\bigg(\sqrt{\frac{\l}{2l_p}}-\l-\frac{(\beta f)^2}{4\l}\bigg),\label{FreeLim}
\end{eqnarray}
where the first term is the contribution from integration over the chain configurations along the two confined axes, and the second is the contribution from the single unconfined axis.  Endpoint effects are neglected in deriving Eq. \ref{FreeLim}, and is only valid for very long chains (where $L$ is larger than all other length scales in the problem).  
The average extension of the chain, under tension, can be calculated using 
\begin{eqnarray}
\langle Z\rangle=\frac{\partial \F}{\partial (\beta f)}=\frac{\beta f L}{2\l(l_p,R,f)}\label{extensionEq},
\end{eqnarray} 
where we have explicitly included the dependence of the mean field solution of $\l$ on the physical parameters at play.  Eq. \ref{extensionEq} is similar to the result found in the case of an unconfined chain under an external tension\cite{MFforce}, which is straightforward to evaluate once the mean field solution for $\l$ is known.  {\textcolor{black}{We note that this result is analytically exact on the Mean Field level, but the value of $\lambda$ determined in this paper neglects the excess endpoint fluctuations.  While these  endpoint effects are important in accurately determining the end-to-end distribution functions\cite{WinklerDist,DistributionReview}, these effects are perturbative for the mean extension \cite{MFforce,MikeTensionPaper} (since eq. 5 is accurate to leading order in $L$ but neglects higher-order corrections).  }
}


To determine the free energy and mean extension, we must solve the MF equations $\partial \F/\partial k=\partial \F/\partial\lambda=0$.  The resulting equations are greatly simplified by noting that 
\begin{eqnarray}
k=\frac{l_pR^2}{2}\w_+^2\w_-^2\qquad &&\qquad \lambda=\frac{l_p}{2}(\w_+^2+\w_-^2)\\
\frac{\partial\w_\pm}{\partial\lambda}=\pm \frac{\w_\pm}{l_p(\w_+^2-w_-^2)}\quad && \quad \frac{\partial\w_\pm}{\partial k}=\mp \frac{1}{l_pR^2\w_\pm(\w_+^2-w_-^2)}.
\end{eqnarray}
After some algebra, the variational equations for $\lambda$ and $k$, respectively, become
\begin{eqnarray}
\frac{1}{l_p(\w_++\w_-)}=1-\frac{1}{\sqrt{8\l l_p}}-\frac{(\beta f)^2}{4\l^2}\quad && \quad\frac{1}{l_p(\w_++\w_-)}=R^2\w_+\w_-. \label{keq0}
\end{eqnarray}
As the left-hand side of both equalities in the above equation are identical, we can readily solve for $k$ in terms of $\lambda$, with,
\begin{eqnarray}
k=\frac{l_p}{2R^2}\bigg(1-\frac{1}{\sqrt{8\l l_p}}-\frac{1}{4}\left(\frac{\beta f}{\l}\right)^2\bigg)^2\label{kSolution}
\end{eqnarray}
so that the confinement parameter $k$ at the mean field level can be determined exactly.  The mean field parameter $\l$, enforcing the inextensibility of the chain, requires the solution of a complicated equation in Eq. \ref{keq0}, after substitution of the exact solution for $k$.  
Although it may not be possible to solve for the exact values analytically for all $l_p$, $R$, and $f$, it is  straightforward to numerically determine the mean field values accurately for any $l_p$ and $R$.   The asymptotic behavior of the roots can be readily determined in certain limits, as discussed in the next section.

\section{Results}

\subsection{Scaling for weak confinement and weak tension}

An asymptotic solution to the Mean Field equations can be determined in a variety of parameter regimes.  The simplest scenario is the limit where both the cylindrical confinement and the external tension are weak.  In the limit of $R/l_f\ll 1$ (with $l_f=k_BT/f$ the Pincus length) and $l_p/R\ll 1$, the chain is weakly confined and we expect the mean field solution to be a perturbation on the solution to the unconfined three dimensional mean field theory.  It has been shown \cite{davePaper} that $\l\sim 9/8l_p$ in the absence of external force or confinement. In order to determine the asymptotic behavior of the mean field parameter, we expand in a series for small values of the rescaled variables $\tilde l_p\equiv l_p/R\ll 1$ and $\varphi=R/l_f\ll 1$, with $\l R =\frac{9}{8\tilde l_p}+\sum_{m,n=0}^\infty b_{nm}{\tilde l_p}{}^n\varphi^m$ for some coefficients $b_{nm}$.  Substitution for the exact solution for $k$ in Eq. \ref{kSolution} into the MF equation for $\lambda$ in eq. \ref{keq0} allows us to perform a series expansion in higher order of $\tilde l_p$ and $\varphi$, both assumed to be small.   Iteratively solving for the lowest order coefficients $b_{ij}$ shows $\lambda \approx \frac{9}{8l_p}-\frac{4l_p}{9R^2}+\frac{4l_p}{9 l_f^2}+O(l_p^3/R^4)$, to leading order in $\tilde l_p$ and $\varphi$.  To leading order, the free energy under weak confinement and weak external tension is
\begin{eqnarray}
\frac{\F}{L}\approx \frac{9}{8l_p}+\frac{2l_p}{9R^2}\left(1-\frac{8l_p^2}{81 R^2}\right)-\frac{2l_p}{3 l_f^2}\qquad (l_p\ll R, R/l_f f\ll 1)\label{WeakAndWeakFree}.
\end{eqnarray}
Note that retaining higher order terms in $\lambda$ does not affect the scaling coefficients in Eq. \ref{WeakAndWeakFree}.  It is interesting to note that the leading order contribution of the tension is independent of the radius of the cylinder, with the confinement only entering into the free energy through coupling with the persistence length.  This is due to the distinct axes over which each energetic contribution acts, each of which are perturbative in this limit.  Not surprisingly for weakly confined chains, the deflection length $l_d=(l_pR^2)^{1/3}$ does not enter into the free energy. We note that these scaling coefficients are not likely to be precise, as has been previously noted for the MF solutions in multiple contexts \cite{MFforce,newMFPaper,MikesPaper}.  However, the scaling with each variable is expected to be accurate.   

For a weakly confined chain, the average extension in eq. \ref{extensionEq} becomes
\begin{eqnarray}
\frac{\langle Z\rangle}{L}\approx \frac{4}{9}\frac{l_p/l_f}{1-{2}\left(\frac{4l_p}{9R}\right)^2}\label{weakWeakExt},
\end{eqnarray}
to fourth order in $\tilde l_p$ and $\varphi$, growing linearly with $f$ as long as $l_p\ll R$.  The linear increase for low forces and weak confinement is shown in the purple triangles of Fig. \ref{extensionFixedR.fig} (A), satisfying the expected linear scaling.

\begin{figure}[h]
\begin{center}
\includegraphics[width=.5\textwidth]{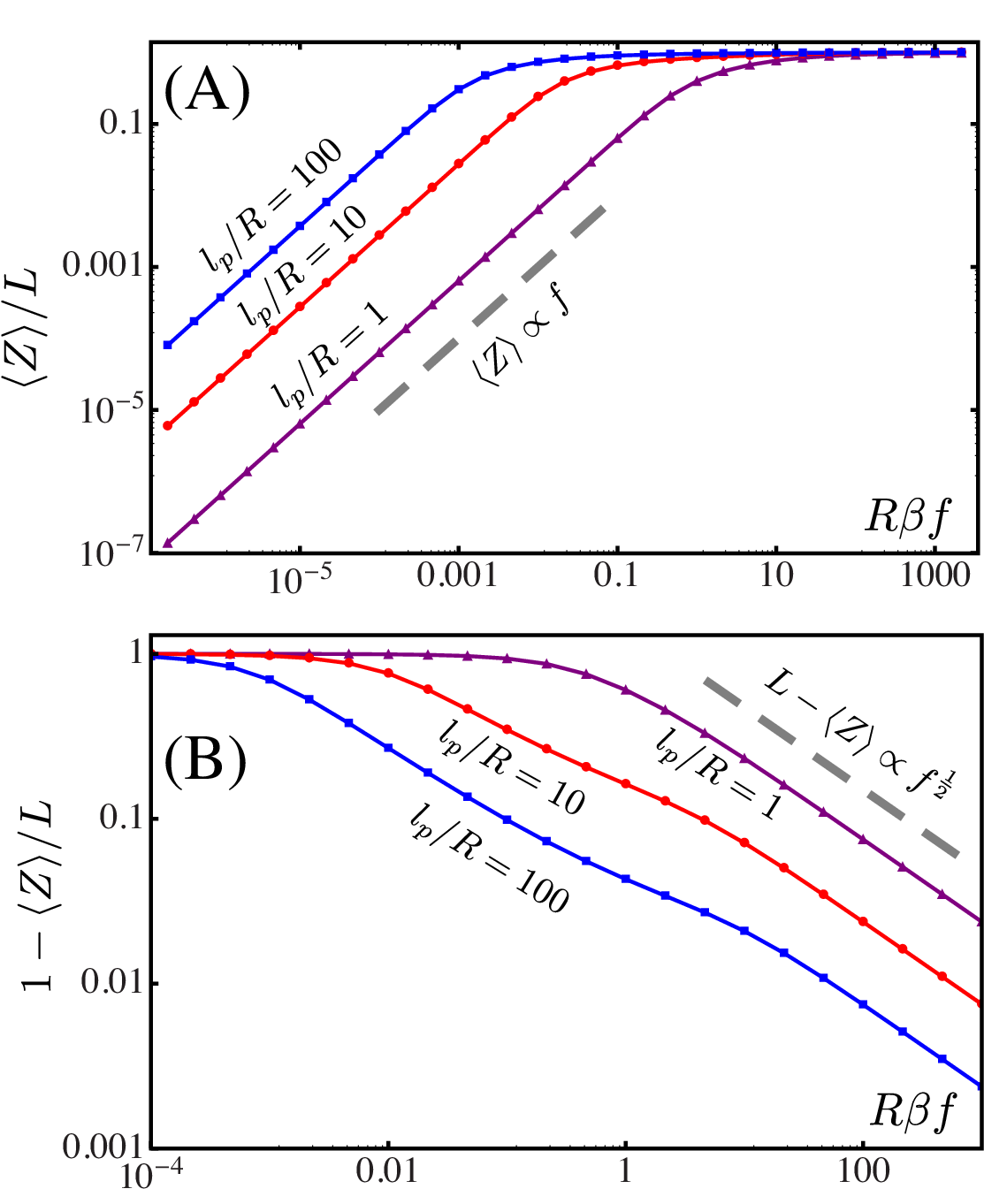}
\caption{ (A) Average chain extension (Eq. \ref{extensionEq}) as a function of $f$ (with $R=1$ held fixed) for various values of $l_p$:   $l_p/R= 1$ (purple triangles), 10 (red circles), and 100 (blue squares).  The extension increases linearly with force for small $f$, and the linear scaling is unperturbed by confinement effects.  (B) The approach to full extension, $1-\langle Z\rangle/L$, as a function of $f$ scales as $f^{-1/2}$ for large external tension, but deviations from this scaling occur for very stiff chains.  {\textcolor{black}{The data are solutions to eq. \ref{FreeLim}, with k given by eq. \ref{kSolution} and with $\lambda$ determined numerically from eq \ref{keq0}.}}}
\label{extensionFixedR.fig}
\end{center}
\end{figure}

\subsection{Strongly confined chains under weak tension}

It is also possible to determine the scaling behavior of the polymer under strong confinement (with $l_p\gg R$) but still constraining the external tension to be weak (with $R\beta f \ll 1$).  A one-dimensional WLC in the absence of tension on the mean field level will satisfy $\lambda \sim 1/8l_p$, and we expect that $\l$ must converge to this value for sufficiently small $R$ or sufficiently large $l_p$ (since transverse fluctuations must vanish in either limit).  The effect of confinement is contained in the mean field variable $k$, which should capture the transverse statistics of the chain.  It is known that the Odijk length scale, $l_d\propto (R^2l_p)^{1/3}$,  emerges for strongly confined WLCs  to the interior of a cylinder \cite{odijk,odijkHarmonic}.  A stiff chain will predominantly be aligned with the cylinder axis, and $l_d$ would be the typical distance along the chain for transverse fluctuations (a consequence of the mean field approach) to encounter the walls (causing a deflection).   A similar scaling has been observed for surface confined chains \cite{Yang07PRE} and we expect the deflection length to emerge at the MF level because the hard constraint is replaced by a soft harmonic potential. As a consequence, the stiff chain `deflects' off the soft harmonic potential rather than a rigid wall.  

In the limit of $\tilde l_p\equiv l_p/R\gg1$ and $\varphi=R/l_f\ll1$ we expect the leading order behavior of $\lambda$ can be recovered with the ansatz $\l\approx \frac{1}{8l_p}+\sum_{n=4}^\infty c_n \tilde l_d^{-n/3}+\sum_{m=1}^\infty d_{m}\varphi^m$.  This is similar to the expression for $\l$ in the weakly confined case but with the expansion in terms of small $\varphi$ and large $\tilde l_p^{1/3}$, and ignores cross-terms (assumed to be higher order, as was the case for weak confinement and low tension).  The restriction of $n\ge 4$ in the sum over $\tilde l_p$ ensures we recover the expected 1-d scaling for $R=0$, since the leading order behavior is expected to be $\lambda=1/8l_p$ for $R=0$.  Substitution into Eqs. \ref{keq0} using \ref{kSolution} and taking the limit of small $\varphi$ and large $\tilde l_p$ shows that $c_{4}=0$ and $c_{5}=1/(4\times 2^{1/3})$ cancel the lowest order terms in $\tilde l_p$, while $d_1=0$ and $d_2=4$ cancels the leading term in $\varphi$.  Substitution of this solution into the free energy 
yields
 \begin{eqnarray}
\frac{\F}{L}& \approx& \frac{3}{2^{5/3}l_d}-\frac{2l_p}{l_f^2}+\frac{l_d}{l_f^2} +\frac{1}{8l_p}\label{StrongAndWeakFree},
\end{eqnarray}
with the first term the leading order in the confinement, the second to leading order in $l_f$, the third term the lowest order coupling between the competing length scales $l_d$ and $l_f$, and the fourth term the 1-dimensional MF free energy.  In the absence of force, this expression for the free energy at a large eternal tension scales as $L/l_d$, agreeing with the known result for a strongly confined WLC \cite{odijk}.  The coefficient of the leading term in Eq. \ref{StrongAndWeakFree} is $3/2^{5/3}\approx 0.945$, which will dominate in the limit of $R\to 0$.  This compares reasonably well with the theoretical prediction for a surface-confined chain, with Eq. 38 in a previous study \cite{Yang07PRE}, which predicts a coefficient of 0.8416 by solving an exact  nonlinear Fokker-Plank equation.  We also note that the coefficient of the leading term of Eq. \ref{StrongAndWeakFree} is exactly twice the value Eq. [8]  in an earlier report of DNA in confined to the interior of a cylinder \cite{odijkHarmonic}. This agreement along with the accurate prediction of the scaling coefficient in the previously studied case of spherical confinement using the mean field theory \cite{Morrison09PRE} suggest that even the calculated numerical coefficients arising  may generally be reliable.   

\begin{figure}[h]
\begin{center}
\includegraphics[width=.5\textwidth]{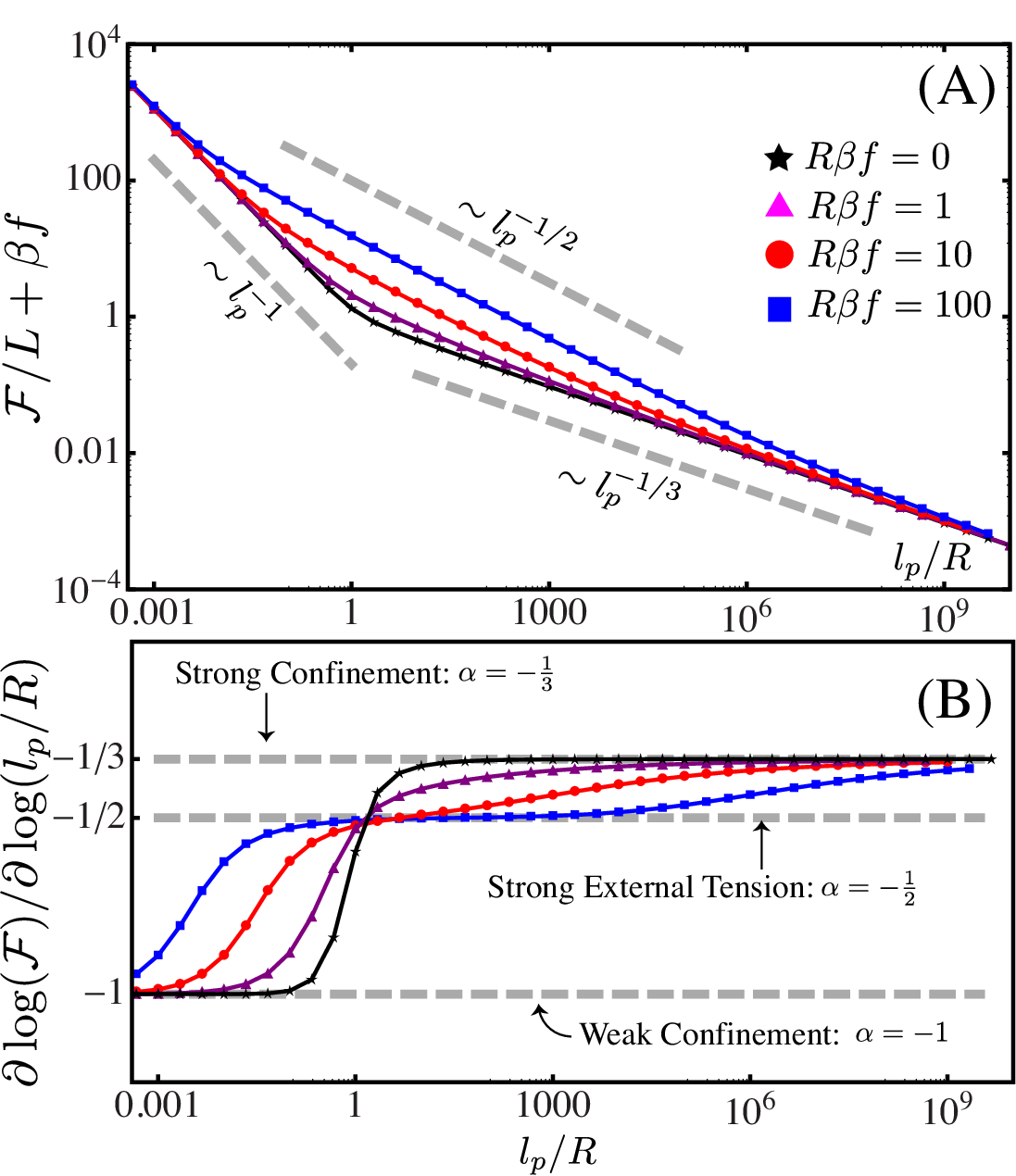}
\caption{(A) Numerical roots for the extensive component of the mean field free energy for $R\beta f=0$ (black stars), 1 (purple triangles), 10 (red circles), and 100 (blue squares) as $l_p$ is varied and $R$ held fixed.  The scalings derived for the three asymptotic regimes are indicated by the gray dashed lines.  The free energy is shifted by a factor $L/l_f$ to counter the leading term in Eq. \ref{strongFreeEnergy}.  (B) The coefficients $\alpha$ for the curves in (A) show that the scaling of $\F\sim l_p^{-1}$ and $\F\sim l_p^{-1/3}$ are recovered for sufficiently weak or sufficiently strong confinement respectively. Furthermore, $\F\sim l_p^{-1/2}$ is observed only for very large forces.  }
\label{asymptoticFree}
\end{center}
\end{figure}

The scaling is confirmed by the numerical calculation for $l_p/R\gtrsim 10$ at $f=0$ in the black stars in Fig. \ref{asymptoticFree}, and for non-zero forces the scaling is recovered for sufficiently strong confinement.  Fig. \ref{asymptoticFree}(B) shows that non-zero tensions can significantly alter the onset of the Odijk regime.  Even an external tension as low as, ${R}/{l_f}=1$ (the purple triangles),  delays the onset of the transition to the Odijk scaling, $\alpha=\partial\log(\F)/\partial\log(l_p/R)=-\frac{1}{3}$, by orders of magnitude in $l_p/R$.  In particular, we note that modeling the histone as a long cylinder of radius $R\approx 3.15$nm and DNA with $l_p\approx50$nm will have a free energy of confinement satisfying the Odijk scaling at $f=0$ (since $l_p/R\approx 16$).   The binding of DNA to histones is known to be interrupted by forces  on the order of $f\approx 10$pN \cite{BenninkNatStructBiol01} and the unbinding of DNA already bound to histones occurs \cite{BenninkNatStructBiol01,Cui00PNAS} at forces $\approx 20-30$pN (or $7.7\lesssim \frac{R}{l_f} \lesssim 23.0$).  
The intermediate external tension in the red squares of Fig. \ref{asymptoticFree} (with ${R}/{l_f}=10$) falls within this range, indicating that the Odijk scaling may not be discernible  by tensile forces over a wide range of biologically relevant conditions.  The calculations suggest that tension effects may be more important than confinement in nucleosomes (see the lower panel in Fig. \ref{asymptoticFree}). 

It is interesting to note that confinement does not have a similarly strong effect on the scaling for the extension of a WLC under tension.  The extension of the chain for small $f$ and large $l_p/R$ becomes
\begin{eqnarray}
\frac{\langle Z\rangle}{L}\approx\frac{4l_p/l_f}{1+\left(\frac{2R}{l_p}\right)^{2/3}},\label{strongWeakExt}
\end{eqnarray}
with the variations in the denominator being weak due to the approximation $l_p/R\gg1$.  Note that the scaling is linear in $l_p$, despite the emergence of the deflection length as the dominant length scale in the free energy in Eq. \ref{StrongAndWeakFree}.  We expect that even stiff chains will deviate only slightly from linear scaling at small to moderate values of $f$ (confirmed in the red squares and blue circles of Fig. \ref{extensionFixedR.fig}(B)).  The leading term in Eq. \ref{strongWeakExt} differs from the extension of a weakly confined chain by a factor of $\frac{1}{9}$, which implies that strongly confined chains are expected to have a greater average extension at the same external tension as weakly confined chains (a physically sensible result).  This is confirmed in Fig. \ref{extensionFixedp.fig}, which shows that a decrease in the radius of the cylinder at fixed $l_p$ decreases the midpoint of extension ($f_m$, the force at which $\langle Z\rangle/L=0.5$) from $l_p\beta f_m\sim 2$ by nearly an order of magnitude. For parameters matching the wrapping of DNA around a histone core, with $l_p\approx 50nm$ and $R\approx3.15$nm, we predict that the midpoint of the transition is  greatly reduced from its unconfined value, with $f_m(R\to \infty)\approx 0.16$pN for the unconfined chain\cite{MFforce} and $f_m(R=3.15)\approx 0.025$pN.  \textcolor{black}{We note that the binding of the DNA to specific sites on the histone are not accounted for in the current theory and may significantly alter the scaling of the free energy.  Binding of DNA to negatively charged colloids\cite{adsorption1,adsorption2} are expected to exhibit an Odijk scaling in the free energy for low tension, but forces $f\gtrsim k_BT/R$ may be sufficient to interrupt the emergence of this scaling.}

\begin{figure}[htbp]
\begin{center}
\includegraphics[width=.5\textwidth]{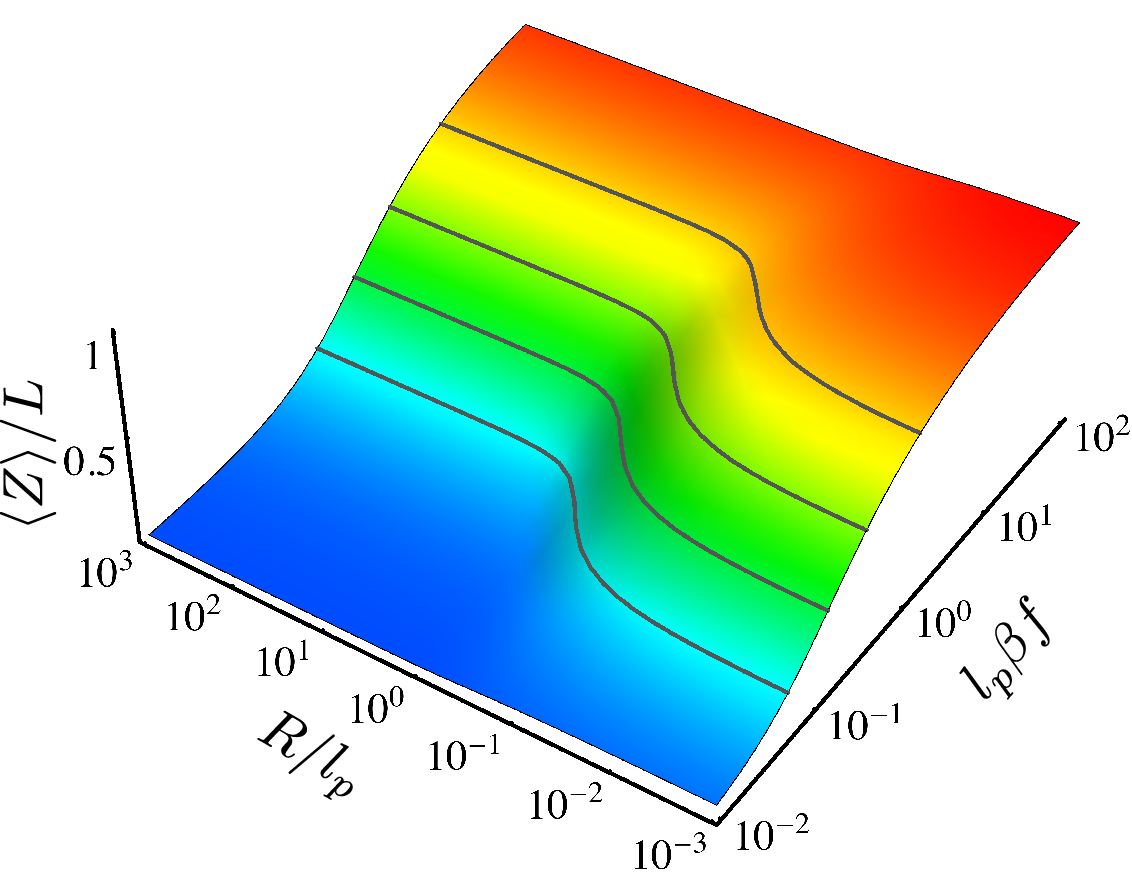}
\caption{Extension $\langle Z\rangle/L$ for fixed $l_p=1$ as $R$ and $f$ are varied.  The midpoint of the force extension curve for sufficiently large $R/l_p\sim 10^3$ occurs near $l_p\beta f_m\sim 2$, but is reduced by nearly a factor of 8.8 for small $R/l_p\sim 10^{-3}$ to $l_p\beta f_{m}\sim 0.23$, in good agreement with the predicted reduction by a factor of 9 between weakly and strongly confined chains. }
\label{extensionFixedp.fig}
\end{center}
\end{figure}

\subsection{Confined chains under high tension}

Finally, we consider the limit of external tensions that are strong enough to be comparable to confinement effects, satisfying $\frac{R}{l_f}\gtrsim l_p/R$.  To determine the asymptotic form of the free energy in this limit, we define $\tilde l_p=l_p/R\equiv \sigma\times R/l_f=\sigma \varphi$ for some unconstrained $\sigma$ (assumed finite, but not restricted to be large or small), and expand $\l$ in the limit of large $\varphi$.  In this limit, it is straightforward to show that $\lambda R\approx \frac{\varphi}{2}+\frac{3}{8\sqrt{\sigma}}+\frac{(9\sigma-8)}{32\sigma \varphi}$ to leading order in $\varphi\gg 1$. 
In the high-force and high-tension limit, the free energy becomes,
\begin{eqnarray}
\frac{\F}{L}&\approx&-\frac{\varphi}{R}+\frac{3}{2R\sqrt{\sigma}}+\frac{1}{2R\varphi}\left(1+\frac{9}{16\sigma}\right)\\
&=& -\beta f+\frac{3}{2}\sqrt{\frac{\beta f}{l_p}}+\frac{9}{32l_p} +\frac{1}{2\beta f R^2}.\label{strongFreeEnergy}
\end{eqnarray}
Note that the free energy here does not depend on the deflection length $l_d\sim (l_p R^2)^{1/3}$ that was seen for strongly confined chains with low tension.  
Thus, a strong force  significantly alters the free energy of confinement.  The predicted scaling is confirmed in the blue circles in Fig. \ref{asymptoticFree} for very large forces (with $R\beta f= 100$).  We do not observe the expected limits of $\F\sim (l_p/l_f)^{-1/2}$ or $\F\sim(l_pR^2)^{-1/3}$ for strong confinement and intermediate tension (the purple triangles or red squares with ${R}/{l_f}=1$ or 10 respectively), but instead an extended crossover region between the expected scaling laws emerges.  The average extension of the chain in this limit is similar to the unconfined WLC, with
\begin{eqnarray}
\frac{\langle Z\rangle}{L}=1-\frac{3}{4\sqrt{l_p\beta f}},\label{highForceF}
\end{eqnarray}
with the highest order dependence on the confinement radius scaling as $({R/l_f})^{-2}$, which is assumed to be small in this limit.  Note that the inextensibility of the chain is respected in the mean field theory (since $\langle Z\rangle \to L$ as $f\to \infty$), and that Eq. \ref{highForceF} (valid for sufficiently high forces) is identical to that of the unconfined chain.  The approach to full extension is shown in Fig. \ref{extensionFixedR.fig}(B), confirming the scaling of $1-\langle Z\rangle/L\sim f^{-1/2}$ for large $f$ regardless of the strength of confinement.  Deviations from this scaling occur only for much weaker forces (with $l_p\beta f\lesssim 1$) for strongly confined chains.

\section{Conclusions}

In this paper, we calculated the free energy and linear extension of a wormlike chain confined to the surface of a cylinder with an applied external tension using a mean field approach, which we showed previously \cite{Morrison09PRE} reproduces the exact results for confinement around the surface of a sphere \cite{Spakowitz03PRL}.  We conclude with the following additional comments. (1) Our method recovers the Odijk scaling of the free energy in the absence of force and the one-dimensional extension profile for a WLC in the limit of small cylinder radius. Previously these relations were derived using sound physical arguments \cite{odijk} and numerical methods \cite{Yang07PRE}.  
(2) The coefficient in Eq. \ref{StrongAndWeakFree} in the free energy expression obtained here is fairly close to the result obtained previously \cite{Yang07PRE}, which shows that the mean field theory not only predicts the scaling relation but also yields the coefficients that are fairly accurate. (3) For a nucleosome, $\frac{l_p}{R}$ lies in the range (5-10) depending on the concentration and valence of counter ions. Moreover, experiments \cite{Cui00PNAS} show that the first stage of DNA unwrapping occurs at $f \approx (3-5)$ pN. For these parameters, the theory predicts that  $\alpha=\partial\log(\F)/\partial\log(l_p/R)\approx -0.5$ (red curve in Fig. \ref{asymptoticFree}B), which corresponds to  external tension effects dominating modest confinement.   

\noindent {\bf Acknowledgements:} DT is grateful to Fyl Pincus for free lessons, with historical connotations, on polymer physics. This work was supported by a grant from the National Science Foundation (CHE 2320256, PHY 2014141 and PHY 2019745) and the Welch Foundation through the Collie-Welch Chair (F-0019). 

\bigskip

{\bf Data Availability Statement:} The data that support the
findings of this study are available from the corresponding
author upon reasonable request.

{\bf Author Contribution Statement:}
GM and DT designed the project, performed calculations, analyzed the data
and wrote the paper.

 \bibliographystyle{unsrt}
 \bibliography{cylinder.bib}

\end{document}